\documentstyle[preprint,revtex,eqsecnum]{aps}
\begin{document}
\draft
\begin{title}
 { Effective Potentials in QCD and\\ Chiral Symmetry Breaking}
 \end{title}
\author{A. Mishra}
\begin{instit}{Physics Department, Utkal University, Bhubaneswar-751004, India}
\end{instit}
 \author {H. Mishra and S. P. Misra}
\begin{instit}
{Institute of Physics, Bhubaneswar-751005, India.}
\end{instit}
\begin{abstract}
We consider chiral symmetry breaking through nontrivial vacuum
structure  with an explicit construct for the vacuum
with quark antiquark condensates in QCD with Coulomb gauge
for different phenomenological potentials.
The dimensional parameter for the condensate function gets
related to $<{\bar \psi }\psi>$ of Shifman, Vainshtein and
Zakharov. We then relate the condensate function
to the wave function of pion as a Goldstone mode. This simultaneously
yields the pion also as a quark antiquark
bound state as a localised zero mode of vacuum.
We then calculate different pionic properties using the wave function
as obtained from the vacuum structure.
\end{abstract}
\pacs{}
\narrowtext

\section {\bf Introduction}
 Vacuum structure in quantum chromodynamics (QCD) is nonperturbative
 and highly nontrivial with quark and gluon condensates in the
 nonperturbative regime\cite{svz}.
Since QCD at low energy is not so far solvable, spectroscopic
properties of hadrons are conventionally dealt with through
effective potentials \cite {cornell,licht2,arp87,lich}.
Also, another attempt was made by
Nambu and Jona-Lasinio (NJL) with
chiral symmetry breaking \cite{njl}
quite sometime back with pion as the Goldstone mode
\cite{mandula,davis,yopr,bhaduri}.
It is clear that, the Goldstone pion is also a quark antiquark bound state.
 Hence through Goldstone theorem pion state
along with its wave function as a quark antiquark pair
should get derived from the vacuum structure.
It is surprising that this particular aspect is absent in the
extensive literature on chiral symmetry breaking leading to
Goldstone pions.

In the present paper we shall link both to obtain
the pion wave function determined from the vacuum structure with
symmetry breaking. For this purpose, we
consider phase transition as a vacuum realignment
with an explicit ansatz, and use the
techniques developed earlier \cite{hm88} to define the gap equation and
Goldstone mode for different potentials.
The gap equation is similar to what is usually derived through
Schwinger Dyson equation for example as in Ref.\cite{davis}.
 As advertised, the other extra result here is that
the pion state as a space localised quark
antiquark zero mode of destabilised vacuum also gets determined.
We also discuss the effects of $approximate$ symmetry breaking where the gap
equation changes giving rise to a change in the pion wave function.
The present dynamical mechanism may also be relevant in the context
of top condensate phenomenology \cite {bardeen,rnm}.

We organise the paper as follows. In section II, we recapitulate
\cite {davis} the vacuum realignment with quark antiqurak pairs
 as a {\em unitary transformation} which includes
an explicit construct for the destabilised
vacuum \cite {hm88} and write the gap equation for any arbitrary potential.
In section III, we then identify the pion as Goldstone mode and
relate its wave function
with functions associated with the vacuum structure \cite {isi}.
We also note the familiar results of current algebra with the present
picture of vacuum realignment.
In section IV we determine the vacuum structure for specific phenomenological
potentials and consider explicitly
exact or approximate chiral symmetry breaking along with the corresponding
pion wave function as a quark antiquark pair.
In section V, we discuss the results.

\section {\bf Vacuum with Quark Condensates}
 We start with the effective Coulomb gauge QCD Hamiltonian
\cite{mandula,davis}
\begin{equation}
{\cal H}(\vec x)=
{\psi(\vec x)^i}^\dagger (-i\vec \alpha \cdot {\vec \bigtriangledown})
{\psi(\vec x)}^i
+\frac{1}{2}\int d\vec y {\psi^i_\alpha}^\dagger(\vec x)
\psi_{\beta}^{j}(\vec x)
V_{\alpha\beta ,\gamma\delta}^{ij,kl}
(\vec x -\vec y){\psi_{\gamma}^k(\vec y)}^\dagger
\psi_\delta^l (\vec y),
\end{equation}
which has chiral invariance.
In the above $i,j$ stand for color indices and
$\alpha,\beta$ stand for the spinor
indices and { $V^{ij,kl}_{\alpha\beta,\gamma\delta} (\vec x -\vec y)$}
 is the potential.
For effective QCD based vector potential we may take
\begin{equation}
V^{ij,kl}_{\alpha\beta,\gamma\delta}(\vec x-\vec y)=
\delta_{\alpha\beta}\delta_{\gamma\delta}
(\frac{\lambda^a}{2})_{ij}
(\frac{\lambda^a}{2})_{kl}
V(\mid \vec x-\vec y\mid),
\end{equation}
where $\lambda^a$ are the  Gellman matrices.
In the above we also have to take the
{\it counter terms} ${\cal H}_C$ given as
\begin{equation}
{\cal H}_C=(Z-1)\bar\psi(\vec x)\vec\gamma.(i\vec\nabla)\psi(\vec x)
\end{equation}
to cancel the divergences \cite{davis}.

The field operators $\psi(\vec x)$ may be expanded as
\begin{equation}
\psi(\vec x)=\frac {1}{(2\pi)^{3/2}}\int \left[U_r(\vec k)c_{Ir}(\vec k)
+V_s(-\vec k)\tilde c_{Is}(-\vec k)\right]e^{i\vec k\cdot \vec x} d\vec k
\end{equation}
where $U$ and $V$ are given by
\begin{equation}
U_r(\vec k)=\frac{1}{\sqrt 2}\left( \begin{array}{c}1\\ \vec \sigma \cdot\hat k
\end{array}\right)u_{Ir} ;\quad V_s(-\vec k)=\frac{1}{\sqrt{2}}\left(
\begin{array}{c}-\vec \sigma .\hat k \\ 1 \end{array}\right)v_{Is}
\end{equation}
for $free$ chiral fields. The perturbative vacuum is defined by this basis
when we have $c_I\mid vac> = 0 $$={\tilde c_I}^{\dagger}\mid vac>$.
We next consider a trial vacuum state given as \cite{hm88}
\begin{mathletters}
\begin{equation}
\mid vac' > = U\mid vac>\equiv exp(B^\dagger - B)\mid vac>
\end{equation}
with
\begin{equation}
B^{\dagger}=\int f(\vec k){c_{Ir}(\vec k)}^\dagger
{u_{Ir}}^\dagger(\vec \sigma .\hat k)
v_{Is}{\tilde c}_{Is}(-\vec k) d\vec k .
\end{equation}
\end{mathletters}
 Here $f(\vec k)$ is a trial function associated as above with quark anti-quark
condensates.
We shall minimise
the energy density of $\mid vac'>$ to analyse the
possibility of phase transition \cite {hm88} from $|vac>$ to $|vac'>$.
For this purpose we first note that with the above transformation the operators
which annihilate $\mid vac'>$ are given as
\begin{equation}
b_I(\vec k)=Uc_{I}(\vec k)U^{-1},
\end{equation}
which with an explicit calculation yields the Bogoliubov transformation
\begin{equation}
\left(
\begin{array}{c} b_{Ir}(\vec k)\\{\tilde b}_{Is}(-\vec k)
\end{array}
\right)
=\left(
\begin{array}{cc}
cosf & -\frac{f}{\mid f\mid}sinf a_{rs}\\
\frac{f^*}{\mid f\mid}sinf (a^\dagger)_{sr} & cosf
\end{array}
\right)
\left(
\begin{array}{c}
c_{Ir}(\vec k)\\
{\tilde c}_{Is}(-\vec k)
\end{array}
\right ).
\label{chi8}
\end{equation}
Here $a_{rs}={u_{Ir}}^{\dagger}(\vec \sigma \cdot \hat k)v_{Is}$.
Using the above transformation (\ref {chi8}) the expectation
value of the Hamiltonian with respect to $\mid vac'>$ is given as
\begin{equation}
{\cal E}=<vac'\mid {\cal H}(x)\mid vac'>\equiv T+V,
\end{equation}
where $T$ and $V$ are the expectation values corresponding to the kinetic
and the potential terms in Eq.(1). With a straightforward evaluation
we then obtain that
\begin{equation}
T=<vac'\mid \psi^i(\vec x)^{\dagger}(-i\alpha . {\vec \bigtriangledown})
\psi^i(\vec x)\mid vac'>
=-\frac{2N}{(2\pi)^3}\int d\vec k\mid \vec k \mid cos 2f(k),
\end{equation}
where $N=N_c\times N_f$ is the total number of quarks.
Similarly the potential term is given as
\begin{equation}
V=\frac{1}{(2\pi)^6}\int {{\tilde V}_{\alpha\beta,\gamma\delta}}^{ij,kl}
(\vec k_1-\vec k_2)
(\Lambda_{+}(\vec k_1))_{\beta\gamma}(\Lambda_-(\vec k_2))_{\delta\alpha}
d \vec k_1 d\vec k_2, \label{v}
\end{equation}
where ${\tilde V}(\vec k)$ is the Fourier transform of the potential
$V(\vec r)$ given as
\begin{equation}
{\tilde V}(\vec k)=\int V(\vec r)e^{i\vec k . \vec r}d \vec k,
\end{equation}
and $\Lambda_{\pm}$ are
\begin{equation}
\Lambda_{\pm}(\vec k)=\frac{1}{2}\left(1\pm \gamma^0 sin2f(k)\pm
\vec \alpha .\hat k cos2f(k)\right).
\end{equation}
The energy density of the nonperturbative
vacuum $|vac'>$ with respect to the perturbative vacuum is then
given as
\begin{eqnarray}
\Delta {\cal E} & = &
<vac'\mid {\cal H} +{\cal H}_C(x)\mid vac'>(f)
-<vac'\mid {\cal H} +{\cal H}_C(x)\mid vac'>(f=0)\nonumber\\
& \equiv & T_1+ (Z-1)T_1 + V_{pot}+ V_{self}
\label{energy}
\end{eqnarray}
where
\begin{eqnarray}
T_1& = & \frac{12\times N_f}{(2\pi)^3}\int d\vec k \sin^2 f(k)\\
V_{pot}& =&
-\frac{2N_f}{(2\pi)^6}\int d\vec {k_1}d \vec {k_2}
{\tilde V}(\mid \vec {k_1}
-\vec {k_2}\mid) [\sin 2f(k_1)\sin 2f(k_2) \nonumber\\& + & 4{\hat {k_1}.\hat
{k_2}}
\sin ^{2} f(k_1) \sin ^{2} f(k_2)]
\end{eqnarray}
\noindent and
\begin{equation}
V_{self} =
\frac{8N_f}{(2\pi)^6}\int d\vec {k_1}d \vec {k_2}
{\tilde V}(\mid \vec {k_1}
-\vec {k_2}\mid) ({\hat {k_1}\cdot\hat {k_2}})
\sin ^{2} f(k_1)
\end{equation}
In the above, $V_{self}$ is divergent for pure Coulomb potential
which is cancelled \cite{davis} by the counter term contribution
$(Z-1)\times T_1$.  For pure
confining potential there is no divergence in $V_{self}$ and the counter term
is zero \cite{davis}. The gap equation can be obtained by minimising
energy functional of equation (\ref{energy}), and is given as
\begin{eqnarray}
|\vec k|sin2f(k)&+&(Z-1)|\vec k|sin2f(\vec k)
= \frac{2}{3}.\frac{1}
{(2\pi)^3}\int d\vec k' \tilde V(|\vec k-\vec k'|)\nonumber \\
& \times &\left[cos2f(k)\;sin2f(k')-\hat k \cdot \hat k' cos2f(k')\;
sin2f(k)\right].\label{gap}
\end{eqnarray}
We note that the above gap equation is the same as has been obtained in
Ref.\cite{davis} with the identification
\begin{equation}
tanf(k)=\psi (k).\end{equation}

\section{\bf Goldstone mode and pion wave function}
 From the gap equation we obtained two solutions for the field operators
corresponding to $sin2f(k)=0$ or $sin2f(k)\not = 0$ along with the
corresponding ground state as $\mid vac>$ or $\mid vac' >$ respectively.
When chiral symmetry remains good,
\begin{equation} {Q_5}^a \mid vac >=0 \end{equation}
where ${Q_5}^a$ is the chiral charge operator given as
\begin{equation}
{Q_5}^a =\int \psi(\vec x)^{\dagger}\frac{\tau^a}{2}\gamma ^5\psi(\vec x)
d \vec x.
\end{equation}
For symmetry broken case however
\begin{equation}
{Q_5}^a\mid vac'>\not = 0
\end{equation}
corresponding to the pion state. To show this we first note that
\begin{equation} \left[{Q_5}^a, H\right]=0
\end{equation}
irrespective of whether ${Q_5}^a$ and $H$ are written in terms of
field operators corresponding to $sin 2f=0$ or $sin 2f\not =0$
since the anticommutation relation between the operators remain unchanged
by the Bogoliubov transformation.
Clearly, for the Goldstone phase, $|vac'>$ is an approximate
eigenstate of $H$ with
${\cal E} V$ as the approximate eigenvalue (V being the total volume).
With $H_{eff}=H-{\cal E} V$, we then obtain from Eq.(23) that
\begin{equation}
H_{eff}{Q_5}^a|vac'>=0
\end{equation}
i.e. the state ${Q_5}^a\mid vac'>$ with zero momentum
has also zero energy corresponding to the massless pion.
Explicitly, using Eq.(3) and Eq.(7), we then obtain
with $q_I$ now  as two component isospin doublet corresponding to
 (u,d) quarks above,
\begin{equation}
\mid \pi^a(\vec 0)>=N_{\pi}\int q_I(\vec k)^\dagger (\frac{\tau^a}{2})
{\tilde q_I}(-\vec k)sin 2f(k) d\vec k \mid vac'>, \label{pi0}
\end{equation}
where, $N_\pi$ is a normalisation constant.
The wave function for pion thus is given as proportional to
$\tilde u(\vec k)\equiv sin2f(k)$.
The isospin and spin indices of $q^\dagger$ and $\tilde q$ for
quarks have been supressed. Further, with
\begin{equation}
<\pi^a(\vec 0)\mid \pi^b(\vec p)>=\delta^{ab}\delta(\vec p),
\end{equation}
the normalisation constant $N_\pi$  is given by
\begin{equation}
{N_{\pi}}^2 \times \frac{N_cN_f}{2}\int sin^2 2f(k) d\vec k =1.
\end{equation}
Clearly the state as in Eq.(~\ref {pi0}) as the Goldstone mode
will be
accurate to the extent we determine the vacuum structure sufficiently
accurately through
variational or any other method so that $|vac'>$ is
 an eigenstate of the Hamiltonian. The above results yield pion
wave function from the vacuum structure for any
example of chiral symmetry breaking, and is the new feature of looking at
phase transition through vacuum realignment \cite{isi} with an explicit
construction.

\subsection{\bf {Approximate chiral symmetry}}
Next we may consider with the present description
the case of adding a small mass term to the
Hamiltonian that breaks the chiral symmetry explicitly. Then ${Q_5}^a
\mid vac'>$ will not be a zero mode and will have finite mass. In fact
the mass of the pion in the lowest order will now be $m_\pi$ formally given as
\begin{equation}
<\pi^a(\vec 0)\mid H_{sb}\mid \pi^a(\vec 0)>=m_{\pi}\delta(\vec 0),
\label{mpi}
\end{equation}
where $H_{sb}$ is the symmetry breaking part of the Hamiltonian
corresponding to the Hamiltonian density
${\cal H}_{sb}=m_0{\bar\psi}\psi$,  $m_0$ being the current quark mass.
The above may be related to $N_\pi$ and pion decay constant as
 follows. Firstly
we note that the identity for pion decay constant is \cite{sakurai}
\begin{equation}
<0\mid {{J_5}^0}^a\mid\pi^a(\vec p)>=\frac{i}{(2\pi)^{3/2}}
\times\frac{c_\pi\times p_0}{\sqrt{2p_0}}\times
e^{i\vec p.\vec x}, \label{sak}
\end{equation}
where, $c_\pi=94 MeV$. The normalisation constant
$N_\pi$ in Eq.(\ref{pi0}) is then given by using
\begin{eqnarray}
{N_\pi}^{-2}\times \delta(\vec 0) & = &<vac'\mid {Q^a}_5
{Q^a}_5\mid vac'>\nonumber\\ & = &
\int <vac'\mid {Q^a}_5\mid \pi^b(\vec p)>d\vec p<\pi^b(\vec p)\mid
{Q^a}_5\mid vac'>,
\end{eqnarray}
where we have saturated the intermediate states with pions.
The index $b$ is summed and there is no summation over the index $a$.
With Eq.(\ref{sak}) and Eq.(30)
 we then have
\begin{equation}
{N_{\pi}}^{-2}=\frac {1}{2}\cdot(2\pi)^3\cdot{m_\pi{c_{\pi}}^2},
\end{equation}
which links $N_{\pi}$ of vacuum structure with pion mass
and pion decay constant.
Then left hand side of the Eq.(\ref{mpi}) is given by, using Eq.(25) and
Eq.(30)
\begin{eqnarray}
& & <\pi^a(\vec 0)\mid H_{sb}\mid\pi^a(\vec 0)> =
\frac{2}{m_\pi {c_\pi}^2}\times\frac{1}{(2\pi)^3}\cdot
<vac'\mid{Q^a}_5 H_{sb} {Q^a}_5\mid vac'>
\nonumber\\& = &
\frac{2}{m_\pi
{c_\pi}^2}\times\frac{1}{(2\pi)^3}\times\frac{1}{2}<vac'\mid\left[\left[{Q^a}_5,H_{sb}
\right], {Q^a}_5\right]\mid vac'>\\ & = &
\frac{2}{m_\pi {c_\pi}^2}\times -\frac{m_0}{2}<vac'|\bar \psi\psi\mid vac'>
\times \delta(\vec 0).
\end{eqnarray}
This yields from Eq.(28) the familiar current algebra result that
\begin{equation}
{m_\pi}^2=-\frac {m_0}{{c_\pi}^2}<\bar \psi\psi>.\label{calg}
\end{equation}

The pion decay constant, $c_{\pi}$ is related to the pion wave
function, $u_{\pi}(\vec k)$ through the relation \cite{spm78}
\begin{equation}
c_\pi=(2\pi)^{-3/2}\times \sqrt \frac {6}{m_\pi}
\int sin 2f(\vec k)u_{\pi}(\vec k)d \vec k,
\end{equation}
\noindent where, as is clear from equations (25) and (27), the normalised
pion wave function, $u_{\pi}(\vec k)$ is given as
\begin{equation}
u_{\pi}(\vec k)=N_{\pi}\sqrt {\frac{N_cN_f}{2}}sin 2f(\vec k).
\label{upi}
 \end{equation}
\noindent Using the equations (27), (35) and (36),
we have the expression for the pion decay constant, $c_\pi$ as
\begin{equation}
c_\pi=\sqrt{\frac {3}{4\pi^{3}m_\pi}}\Bigl (\int sin ^{2} 2f(\vec k)
d \vec k\Bigr )^{1/2},
\end{equation}
\noindent which is then calculated for different
phenomenological potentials.

 We now note that the gap equation will change for the case of approximate
 chiral symmetry breaking. This will be obtained through the minimisation
 of the modified energy functional given as
\begin{eqnarray}
\Delta {\cal E} & = &
<vac'\mid {\cal H} +{\cal H}_C(x)\mid vac'>(f)
-<vac'\mid {\cal H} +{\cal H}_C(x)\mid vac'>(f=0)\nonumber\\
& \equiv & T_1+ (Z-1)T_1 + V_{pot}+ V_{self}+V_{sb}
\label{energysb}
\end{eqnarray}
where the symmetry breaking contribution is given as
\begin{equation}
V_{sb}=-\frac{3}{\pi^2}m_0 N_f\int dk k^2 sin2f(k).
\label{sbpt}
\end{equation}
Minimisation of the energy functional Eq.(\ref{energysb}) will naturally change
the gap equation making it dependant on $m_0$ through
$V_{sb}$, leading to a change in the variational function $f(k)$ for
the vacuum structure. This modified vacuum structure in its turn will
again lead to the pion wave function
as an {\it approximate} Goldstone mode through Eq.(\ref{upi}).

\subsection{\bf Charge radius of pion}
With the wave function of the pion as above, we may estimate the size of the
Goldstone pion as related to the length scales of vacuum realignment.
The pion state
with momentum $\vec p$ using translational invariance from Eq.(25) becomes
\begin{equation}
|\pi^+(\vec p)>= {N_\pi}\int d\vec k {q_I}^i(\vec k+
\frac{\vec p}{2})^\dagger (\tau^+)_{ij}{\tilde q_I}^j
(-\vec k+\frac{\vec p}{2})
\tilde u(\vec k) |vac'>.
\end{equation}
In Breit frame the electric form factor is given by \cite{spm78}
\begin{equation}
G_E(t)=(2\pi)^3<\pi^+(-\vec p)|J_0|\pi^+(\vec p)>
\end{equation}
where $t=-4p^2$ and $J_0=e\psi^\dagger\psi$.
This may be evaluated directly as
\begin{eqnarray}
G_E(t)&=& e {N_\pi}^2\times
\int d\vec k \tilde u(\vec k-\frac{\vec p
}{2})^*\tilde u(\vec k +\frac{\vec p}{2})\nonumber \\
& \times & \left\{u_1(\vec k-\vec p)u_1(\vec k
+\vec p)+(k^2-p^2)u_2(\vec k -\vec p)u_2(\vec k+\vec p)\right\}.
\label{get}
\end{eqnarray}
In the above
\begin{equation}
u_1(\vec k)=\sqrt{\frac{(1+sin2f
(|\vec k|))}{2}} ;\qquad u_2(\vec k)=\frac{1}{ |\vec k|}\sqrt{\frac{
(1-sin2f(|\vec k|))}{2}}.
\end{equation}
To calculate the charge radius we expand the above in powers of $\vec p$
and the coefficient of $p^2$ will be related to the charge radius through
\begin{equation}
G_E(t)=e(1+\frac{1}{6}R_{ch}^2 t +\cdots).
\end{equation}
With $G_E(t)$ as in Eq(\ref{get}) we then obtain that
\begin{eqnarray}
<R_{ch}^2>& = & \frac{1}{2}\int d\vec k \bigg[\frac{1}{4}({u_0'(k)}^2-
\frac{2}{k}u_0'u_0-u_0''u_0)\nonumber\\
&  & +u_0^2\left\{(u_1'^2-\frac{2}{k}u_1'u_1-u_1''u_1)+
3u_2^2+k^2(u_2'^2-\frac {2}{k}u_2'u_2-u_2''u_2)\right\}\bigg],
\end{eqnarray}
where we have substituted
\begin{equation}
u_0(k)=\frac{1}{\sqrt{(\int sin^2 2f(k) d\vec k)}}\times \tilde u(k),
\end{equation}
and, primes denote differention with respect to $k$.

The above formula applies for any known vacuum
realignment with condensates.
Let us now estimate the charge
 radius with different forms of the potential.
We shall also use Eq.(34) for pion decay constant so that chiral symmetry
was approximately true.
With ${\cal H}_{sb}= m_{0}\bar\psi\psi$, the extra contribution
to the energy density is $m_{0}\times
\left[<\bar\psi \psi>_{vac'}-<\bar  \psi \psi>_{vac}\right]$

We thus consider here chiral symmetry breaking
as a vacuum realignment with an explicit construct for destabilised vacuum.
The new feature of this approach \cite{hm88,isi} is that it enables us to
relate the function that describes
the vacuum structure to the wave function of the localised
Goldstone mode. This language is not only physically appealing
reproducing the conventional results but also
puts severe constraints
on the parameters for symmetry breaking as
illustrated here with different forms of potentials.

\section{\bf Effective phenomenological potentials}
We shall now consider the cases for different
potentials explicitly. For this purpose the condensate
function $f(\vec k)$ of Eq.(\ref{energy}) or of Eq.(\ref{energysb})
should get determined through
energy minimisation. However, since the determination of $f(\vec k)$ is
not possible through functional differentiation analytically,
we shall consider an alternative choice of taking a suitable ansatz
function, the parameters of which will be determined through
numerical minimisation of the  energy density. Since condensation is a
large distance effect we take the ansatz
\begin{equation}
tan 2f(k)= \frac{A\;exp (-B k^2)}{|\vec k|},
\label{ansatz}
\end{equation}
so that for large momentum the condensate function vanishes
as may be expected. Further, the above ansatz is also motivated
by the fact that for free massive fermions
\begin{equation}tan 2f(k)=\frac{m}{|\vec k|}.
\end{equation}
Thus the small $\vec k$ behaviour is correctly simulated through the
ansatz above, with, $``A"$ to be identified as the dynamically generated
mass at zero momentum arising from chiral symmetry breaking. We may
in fact then define the momentum dependant mass as
\begin{equation}
M(k)=A \exp(-Bk^2)
\end{equation}
corresponding to the pole of the propagator of Ref. \cite{davis}.
We shall thus proceed to obtain the results through a simple ansatz
as in Eq.(\ref{ansatz}) instead of using the more sophisticated
over relaxed Gauss-Seidel algorithm as in Ref. \cite{davis}.

Now we shall consider different potentials explicitly.

\subsection {\bf Coulomb potential}
The pure Coulomb potential is given by
\begin{equation}
V(r)=\frac{\alpha}{r}
\end{equation}
In this case the expression for the {\em renormalised}
energy density is given as
\begin{eqnarray}
{\cal E}_R & = &  \frac {6N_f}{\pi^2}\int dk k^3 \sin ^2 f(k)
-\frac{2N_f}{(2\pi)^6}\int d\vec {k_1}d \vec {k_2}
\frac{4\pi\alpha}{(|\vec k_1-\vec k_2|)^2}
 [\sin 2f(k_1)\sin 2f(k_2) \nonumber\\& + & 4{\hat {k_1}.\hat {k_2}}
\sin ^{2} f(k_1) \sin ^{2} f(k_2)]+V_{sb}
\label{chi21}
\end{eqnarray}
and $V_{self}$ cancels with the counter term \cite{davis}.
To determine the parameters A and B  of the condensate function, we
first write down the energy density in dimensionless
units. Thus, with $A'=A\sqrt B$, $x= k\sqrt B$ and $\mu_0=m_0 \sqrt{B}$
the expression for ${\cal E}_R$ becomes
\begin{equation}
{\cal E}_R=\frac{1}{B^2}\times F(A')
\end{equation}
where,
\begin{eqnarray}
F(A') & = & \frac{3N_f}{\pi^2}\int dx x^3
\Bigl (1-\frac{x}{(x^2+A'^2 \exp(-2x^2))^{1/2}}\Bigr )
 +  \frac{\alpha N_f}{\pi^3}\int xdxx'dx'\nonumber\\
&&\Bigg [\Bigl (1-\frac {x}{(x^2+A'^{2}\exp(-2x^{2}))^{1/2}}\Bigr )
\Bigl (1-\frac {x'}{(x'^2 +A'^{2}\exp(-2x'^2))^{1/2}}\Bigr ) \nonumber\\
  - &&   \Bigl \{ \frac {A'^{2}\exp(-(x^{2}+x'^{2}))}
{{(x^ 2+A'^{2}\exp(-2x^{2}))^{1/2}}
{(x' ^ 2+A'^{2}\exp(-2x'^{2}))^{1/2}}}
+\Bigl (\frac {x^2+x'^2}{2xx'}\Bigr )\nonumber\\
 \times && \Bigl (1-\frac {x}{(x ^2+A'^{2}\exp(-2x^{2}))^{1/2}}\Bigr )
\Bigl (1-\frac {x'}{(x' ^2+A'^{2}\exp(-2x'^{2}))^{1/2}}\Bigr ) \Bigr \}
ln \Big | \frac {x+x'}{x-x'}\Big | \Bigg ]\nonumber\\
-&& \frac{3\mu_0}{\pi^2}\int dx x^2 \frac{A'\exp(-x^2)}{(x^2+A'^2
\exp(-2x^2))^{1/2}}.
\end{eqnarray}
\noindent The first and second terms correspond to the kinetic and
potential terms respectively, and the third term corresponds to the
symmetry breaking term. The function $F(A')$ is then
extremised over $A'$. The Coulomb potential has no scale parameter. We may
however determine the scale parameter $B$ from the
SVZ parameter \cite{svz}, which for $N_f=2$, is
\begin{equation}
<\bar\psi \psi>=-\frac{0.55}{4\pi^2} GeV^3
\label{SVZ}
\end{equation}
 If $|vac'>$ after extremisation described the physical vacuum in QCD
then
\begin{eqnarray}
<:\bar\psi \psi:>_{A=A_{min}}&=&
\left[<\bar\psi \psi>_{vac'}-<\bar  \psi \psi>_{vac}\right]_{A=A_{min}}
\nonumber\\
&=&-6N_f\times (2\pi)^{-3}\int d\vec k sin 2 f(k)\Big |_{A=A_{min}}.
\label{svz1}
\end{eqnarray}
\noindent A non-zero value for $<:{\bar \psi }\psi :>_{vac'}$
implies the spontaneous breaking of chiral symmetry.

We note that for energy minimisation $A'_{min}$ vanishes when $\alpha <
\alpha _c=1.28$, and for $\alpha>\alpha _c$ is nonzero with ${\cal E}_R$
as negative, indicating breaking of chiral symmetry. In this case we determine
value of the scale parameter $B$ from the SVZ parameter using equations
(\ref{SVZ}) and (\ref{svz1}). With  $A_{min}={\sqrt B}A'_{min}=M(0)$, we
plot the dynamical quark mass as curve I for Fig.1 and vacuum energy density
${\cal E}_R$ as curve II of Fig.1 against the coupling parameter $\alpha$.
 The ansatz function
$f(\vec k)$ includes a correlation length proportional to
$\sqrt {B}$ for quark condensates, which is plotted as curve III in
Fig.1. This length scale appears to be of the order of
one fermi, which looks reasonable in the context of confinement.
All the curves of fig.1 show that $\alpha_{c} \simeq 1.28$.
This is parallel to the results of Miransky \cite{miransky} where the
critical coupling was 1.05.
  The condensate function $f(k)$ for this case is also plotted in curves
  I, II and III of Fig.2a for coupling $\alpha=1.7$ and for current quark
masses
  $m_0$= 0, 5 and 10 MeV respectively as functions of $k$.
In Fig.2b, we plot the momentum dependant mass-like function $M(k)$ of
equation (49) for $m_0$ = 0, 5 and 10 MeV for the same coupling
against $k$.
We also plot in Fig.2c the corresponding pion wave functions as
(aproximate) Goldstone modes in curves I, II, III respectively, this
last result being a conclusion for the present way of looking at
symmetry breaking.
With these parameters we obtain the dynamical quark mass as 348 MeV,
$f_{\pi}$ as 139 MeV and the square of pion charge radius $R_{ch}^2$
as .64 fm$^2$ for $m_0=0$. All the above parameters increase with
increase in coupling strength.

The calculations for this potential, which is unrealistic for the
light quark sector, is only meant to be an illustration of the
methodology.

\subsection {\bf Pure Confining Potential}
 We shall here consider the case of a linear confining potential given as
\begin{equation}
V(r)=-Cr
\end{equation}
We then have the expression for  the renormalised energy density
given as
\begin{equation}
{\cal E}_{R}=C^2 \left[\epsilon_{F}+\epsilon _{int}^{pot}
+\epsilon _{int}^{self}+{\epsilon _{sb}}\right]
\end{equation}
\noindent where
\begin{equation}
\epsilon_{F} =\frac{3N_f}{\pi^2}\int dx x^3
\left [1-\frac{x}{(x^2+A'^2exp(-2bx^2)) ^{1/2}}\right ]
\end{equation}
\begin{equation}
\epsilon _{int}^{pot}=32\pi\int dr' r'^{3} [F_{1}(r')^2
+{F_2}^{'}(r')^{2}],
\end{equation}
\noindent with
\begin{equation}
F_{1}(r')=\frac{1}{4\pi^2}\int dx x
\frac{A'exp(-bx^2)}{(x^2+A'^2exp(-2bx^2)) ^{1/2}}
\frac{\sin (xr')}{r'},
\label{chi30}
\end{equation}
\begin{equation}
 {F_2}^{'}(r')=\frac{1}{4\pi^2}\int d x
\left [1-\frac{x}{(x^2+A'^2exp(-2bx^2)) ^{1/2}}\right ]
\Bigl (\frac{x\cos (xr')}{r'}-
\frac{\sin (xr')}{r'^2}\Bigr ),
\label{chi31}
\end{equation}
\noindent and
\begin{equation}
\epsilon ^{self}_{int}=-\frac{8N_f}{\pi^3}\int  dx x
\left [1-\frac{x}{(x^2+A'^2 \exp (-2bx^2))^{1/2}}\right ].
\end{equation}
Further, the symmetry breaking contribution is given as
\begin{equation}
{\epsilon }_{sb}=
-\frac{3\mu_0}{\pi^2}\int dx x^2 \frac{A'\exp(-bx^2)}{(x^2+A'^2
\exp(-2bx^2))^{1/2}}.
\end{equation}
\noindent In the above $A'=A/\sqrt{C}$, $b=B\times C$, $r'=r\times
\sqrt{C}$, $x=k/\sqrt{C}$ and $\mu_0=m_0/\sqrt{C}$.

The energy density is now varied with respect to the two parameters $A'$
and $b$.
  The condensate function $f(k)$ for this case as earlier is plotted in curves
  I, II and II of Fig.3a for coupling for $\sqrt C$ = 300 MeV of Adler and
Davis \cite {davis} for current quark masses
$m_0$= 0, 5 and 10 MeV respectively. The curves are not sensitive to these
masses. In Fig.3b, we plot the corresponding curves for $M(k)$ of
equation (49), and, in Fig.3c, pion wave functions as
(aproximate) Goldstone modes.
With these parameters we obtain the dynamical quark mass as 58 MeV,
$f_{\pi}$ as 61 MeV, the square of pion charge radius $R_{ch}^2$
as 3.8 fm$^2$ for $m_0=0$, and the SVZ parameter $-<\bar \psi \psi>=
(98 MeV)^3$. The above values broadly agree with the calculated results
of Adler and Davis \cite{davis} as well as indicate the limitations of
taking a simplified ansatz function. We have also calculated the above
quantities for the parameters of Alkofer and Amundsen \cite{davis} with
our results mostly in agreement with their calculations. We may however
note that the calculated values are {\it not} in agreement with the
experimental values in either case, and the present methodology while
obtaining the pion wave function further confirms this disagreement. Our
objective for taking this potential has been to compare our method
with that of Ref. \cite{davis}.

\subsection{\bf Cornell Potential}
We now consider a more realistic potential given as \cite{lich}
\begin{equation}
V(r)\equiv -\frac {3}{4}V_c(r)=
-\frac {3}{4}\times (-\frac {C_1}{r} + C_2 r +C_3)
\end{equation}
We then have the expression for  the renormalised energy density as
\begin{equation}
{\cal E}_{R}={C_2}^2 \left[\epsilon_{F}+\epsilon _{int}^{pot}
+\epsilon _{int}^{self}+\epsilon_{sb}\right]
\end{equation}
\noindent where
\begin{equation}
\epsilon_F=\frac{3N_f}{\pi^2}\int dx x^3
\left [1-\frac{x}{(x^2+A'^2exp(-2bx^2)) ^{1/2}}\right ]
\end{equation}
\begin{equation}
\epsilon _{int}^{pot}=24\pi\int dr' r'^{2}(r'-\frac {C_1}{r'}) [F_{1}(r')^2
+{F_2}^{'}(r')^{2}],
\end{equation}
\noindent with $F_1(r')$ and $F_2(r')$ given through Eq.(\ref{chi30}) and
Eq.(\ref{chi31}) respectively and
\begin{equation}
\epsilon ^{self}_{int}=-\frac{6N_f}{\pi^3}\int  dx x
\left [1-\frac{x}{(x^2+A'^2exp(-2bx^2)) ^{1/2}}\right ]
\end{equation}
Also, the symmetry breaking contribution is given as
\begin{equation}
{\epsilon }_{sb}=
-\frac{3\mu_0}{\pi^2}\int dx x^2 \frac{A'exp(-bx^2)}{(x^2+A'^2
exp(-2bx^2))^{1/2}}.
\end{equation}
\noindent In the above $A'=A/\sqrt{C_2}$, $b=B\times C_2$, $r'=r\times
\sqrt {C_2}$ and $\mu_0=m_0/ \sqrt {C_2}$.
We may note that the constant term $C_3$ in the potential
does not contribute to the energy density. We take the constants $C_1$ and
$C_2$
respectively as 0.47 and 0.186 GeV$^2$ \cite{lich}.
  The condensate function $f(k)$ for this case is also plotted in curves
  I, II and III of Fig.4a for the parameters as above,
  and for current quark masses $m_0$= 0, 5 and 10 MeV respectively.
In Fig.4b, we plot $M(k)$ of
equation (49), and, in Fig.4c, the corresponding pion wave functions
(aproximate) Goldstone modes.
With these parameters we obtain the dynamical quark mass as 84 MeV,
$f_{\pi}$ as 42 MeV, the square of pion charge radius $R_{ch}^2$
as 2 fm$^2$ and $-<\bar\psi \psi>=(132 MeV)^3$ for $m_0=0$.
All the above quantities increase with current quark mass, except the
charge radius, which decreases. For example, for $m_0=10 MeV$, we get the
dynamical quark mass as 96 MeV, $f_{\pi}=64 MeV$, $R_{ch}^2$ = 1 fm$^2$,
and $-<\bar\psi\psi>=(201 MeV)^3$. We thus find that the increase is
not commensurate with the experimental numbers, which may be attributed
to the fact that the potential is adjusted for bottomonium spectroscopy,
or, to the inadquacy of the potential picture as such.

\section{\bf Discussions}
Let us briefly recapitulate the results here. Firstly, we have
considered phase transition as a vacuum realignment with an
explicit construction for any effective potential through the minimisation
of energy density, and then obtained the pion as an approximate zero mode
of destabilised vacuum.  We also verify that
the present method is in general agreement with the earlier results
of proceeding through Schwinger Dyson equations, and current algebra.
In addition, we  have more constraints
since the pion wave function as a quark antiquark state also gets
determined through the vacuum structure for breaking of chiral symmetry,
and a Schrodinger like wave function for the pion is obtained as an
output without considering any parallel equations of spectroscopy.
The technology thus
gives a physically more appealing picture for phase transition as well
for the pion as a Goldstone mode for an effective potential, and helps
us to understand in a quantitative manner the special role of the pion
for hadron spectroscopy.

However, when we compare the results with experimental numbers,
we conclude that the condensate structure with chiral symmetry
breaking obtained from phenomenological effective potentials are
numerically not consistent with the parameters for spectroscopy.
This implies that chiral symmetry breaking does not most likely come
through effective potentials, but independently, and further illustrates
that the idea of the effective potentials can only have limited validity.
We should look beyond it for the microscopic physics it might hide, and
for this purpose gluon condensates \cite{svz,qcdt0} may be important
and indispensable.

\acknowledgements
The authors are thankful to Snigdha Mishra, S. N. Nayak and
P. K. Panda for many useful discussions. SPM acknowledges to Department of
Science and Technology, Government of India for the research grant
SP/S2/K-45/89 for financial assistance and AM would like to acknowledge
the Council of Scientific and Industrial Research (C. S. I. R.)
 for a fellowship.

\newpage
\centerline{\bf Figure Captions}
\bigskip
\noindent {\bf Fig.1:} In curves I, II and III we respectively plot $A_{min}$
in units of 200 MeV, ${\cal E}_R$ in units of 10 MeV/fm$^3$ and $\sqrt B$
in units of 0.3 fms against $\alpha$ for the Coulomb potential.\hfil
\medskip

\noindent {\bf Fig.2a:}  We plot
the condensate function $f(k)$ against $k$ in GeV for current quark masses
0, 5 and 10 MeV for Coulomb potential as solid,
dashed and dot-dashed curves respectively.\hfil
\medskip

\noindent {\bf Fig.2b:} We plot
the dynamical quark mass $M(k)$ in GeV against $k$ in GeV for current
quark masses 0, 5 and 10 MeV for Coulomb potential as solid,
dashed and dot-dashed curves respectively.\hfil
\medskip

\noindent {\bf Fig.2c:} We plot
the pion wave function $u_{\pi}(k)$ in GeV$^{-3/2}$ against $k$ in GeV
for current quark masses 0, 5 and 10 MeV for Coulomb potential as solid,
dashed and dot-dashed curves respectively.\hfil
\medskip

\noindent {\bf Fig.3a:} We plot
the condensate function $f(k)$ against $k$ in GeV for current quark masses
0, 5 and 10 MeV for confining potential as solid,
dashed and dot-dashed curves respectively.\hfil
\medskip

\noindent {\bf Fig.3b:} We plot
the dynamical quark mass $M(k)$ in GeV against $k$ in GeV for current
quark masses 0, 5 and 10 MeV for confining potential as solid,
dashed and dot-dashed curves respectively.\hfil
\medskip

\noindent {\bf Fig.3c:} We plot
the pion wave function $u_{\pi}(k)$ in GeV$^{-3/2}$ against $k$ in GeV
for current quark masses 0, 5 and 10 MeV for confining potential as solid,
dashed and dot-dashed curves respectively.\hfil
\medskip

\noindent {\bf Fig.4a:} We plot
the condensate function $f(k)$ against $k$ in GeV for current quark masses
0, 5 and 10 MeV for Cornell potential as solid,
dashed and dot-dashed curves respectively.\hfil
\medskip

\noindent {\bf Fig.4b:} We plot
the dynamical quark mass $M(k)$ in GeV against $k$ in GeV for current
quark masses 0, 5 and 10 MeV for Cornell potential as solid,
dashed and dot-dashed curves respectively.\hfil
\medskip

\noindent {\bf Fig.4c:} We plot
the pion wave function $u_{\pi}(k)$ in GeV$^{-3/2}$ against $k$ in GeV
for current quark masses 0, 5 and 10 MeV for Cornell potential as solid,
dashed and dot-dashed curves respectively.\hfil
\newpage
{\bf
f(k)$\longrightarrow$\hfil f(k)$ \longrightarrow$\hfil
\bigskip

f(k)$ \longrightarrow$\hfil f(k)$ \longrightarrow$\hfil
\bigskip

 f(k)$ \longrightarrow$\hfil f(k)$ \longrightarrow$\hfil
\bigskip

 f(k)$ \longrightarrow$\hfil f(k)$ \longrightarrow$\hfil
\bigskip

 u$_{\pi}$(k) in GeV$^{-3/2}$ $ \longrightarrow$\hfil M(k) in GeV$
\longrightarrow$\hfil
\bigskip

 u$_{\pi}$(k) in GeV$^{-3/2}$ $ \longrightarrow$\hfil M(k) in GeV$
\longrightarrow$\hfil
\bigskip

 u$_{\pi}$(k) in GeV$^{-3/2}$ $ \longrightarrow$\hfil M(k) in GeV$
\longrightarrow$\hfil
\bigskip

 u$_{\pi}$(k) in GeV$^{-3/2}$ $ \longrightarrow$\hfil M(k) in GeV$
\longrightarrow$\hfil
\bigskip

 u$_{\pi}$(k) in GeV$^{-3/2}$ $ \longrightarrow$\hfil M(k) in GeV$
\longrightarrow$\hfil
\bigskip

 u$_{\pi}$(k) in GeV$^{-3/2}$ $ \longrightarrow$\hfil M(k) in GeV$
\longrightarrow$\hfil
\bigskip

 k in GeV $\longrightarrow$\hfil k in GeV $\longrightarrow$\hfil
\bigskip

 k in GeV $\longrightarrow$\hfil k in GeV $\longrightarrow$\hfil
\bigskip

 k in GeV $\longrightarrow$\hfil k in GeV $\longrightarrow$\hfil
\bigskip

 k in GeV $\longrightarrow$\hfil k in GeV $\longrightarrow$\hfil
\bigskip

 k in GeV $\longrightarrow$\hfil k in GeV $\longrightarrow$\hfil
\bigskip

 k in GeV $\longrightarrow$\hfil k in GeV $\longrightarrow$\hfil
\bigskip

 k in GeV $\longrightarrow$\hfil k in GeV $\longrightarrow$\hfil
\bigskip

 k in GeV $\longrightarrow$\hfil k in GeV $\longrightarrow$\hfil}

\begin{references}
 \bibitem {svz} M.A. Shifman, A.I. Vainshtein and V.I. Zakharov,
 Nucl. Phys. B147, 385, 448 and 519 (1979);
 R.A. Bertlmann, Acta Physica Austriaca 53, 305 (1981).
\bibitem {cornell} E. Eichten, K. Gottfried, T. Kinoshita, J. Kogut,
K. D. Lane and T. M. Yan, Phys. Rev. Lett. 34, 369 (1975).
\bibitem {licht2} G. Fogleman, D. B. Lichtenberg and J. G. Wills,
Lett. Nuovo Cimento 26, 369 (1979).
\bibitem {arp87} X. T. Song and H. Lin, Z. Phys. C34, 223 (1987);
S.P. Misra, S. Naik and A.R. Panda, Pramana (J. Phys.) 28, 131 (1987).
\bibitem{lich}
D.B. Litchenberg, E. Predazzi, R. Roncaglia, M. Rosso and J. G. Wills,
Z. Phys. C41, 615 (1989).
\bibitem{njl} Y. Nambu and G. Jona-Lasinio, Phys. Rev. 122, 345 (1961);
124, 246 (1961).
\bibitem{mandula} J.R. Finger and J.E. Mandula, Nucl.Phys.B199,
168 (1982).
\bibitem{davis}
 S.L. Adler and A.C. Davis,
Nucl. Phys. B244, 469 (1984); R. Alkofer and P. A. Amundsen,
Nucl. Phys.B306, 305 (1988); A.C. Davis and A.M. Matheson DAMTP 91-34 (1991).
\bibitem{yopr}
A. Amer, A. Le Yaouanc, L. Oliver, O. Pene and
J.C. Raynal, Phys. Rev. Lett. 50, 87 (1983);
ibid, Phys. Rev. D28, 1530 (1983).
\bibitem{bhaduri}S. Li, R. S. Bhalerao and R. K. Bhaduri, Int. J. Mod. Phys.
A6, 501 (1991);
V. Bernard, Phys. Rev. D34, 1601 (1986).
\bibitem {hm88} H. Mishra, S.P. Misra and A. Mishra,
Int. J. Mod. Phys. A3, 2331 (1988);
A. Mishra, H. Mishra, S. P. Misra and
S. N. Nayak, Phys. Lett.251B, 541 (1990); ibid,
Phys. Rev. D44, 110 (1991); Pramana (J. Phys.) 37, 59 (1991).
\bibitem{bardeen}
 Y. Nambu in New Theories in Physics, Proceedings of
the Eleventh International Symposium on Elementary Particle Physics, Kazimierz,
Poland, edited by Z. Ajduk, S. Pokoroski and A. Trautman (World Scientific,
Singapore, 1989), W. J. Marciano, Phys. Rev. Lett. 62, 2793 (1989); Phys.
Rev. D41, 219 (1990).
W. A. Bardeen, C. T. Hill and M. Linder,
Phys. Rev. D41, 1647 (1990).
\bibitem{rnm}K.S. Babu and Rabindra N. Mohapatra, Phys. Rev. Lett. 66, 556
(1991).
\bibitem{isi}S. P. Misra, Talk at Symposium on Quantum Field Theory
and Statistical Mechanics, Calcutta, 1992, (to appear in the proceedings).
\bibitem{sakurai}J.J. Sakurai in $``$Currents and Mesons", The university of
Chicago press 1969 p.83.
\bibitem{spm78}S. P. Misra, Phys. Rev. D18, 1661, 1673 (1978). We note that
$f(\vec k)$ and $|\vec k| g(\vec k )$ of Ref. \cite{spm78} are $sin 2f(k)$
and $cos 2f(k)$ respecitively.
\bibitem{miransky} V.A. Miransky, Nuov. Cim. 90A, 149 (1985).
\bibitem{qcdt0} A. Mishra, H. Mishra, S. P. Misra and
S. N. Nayak,  Pramana (J. Phys.) 37, 59 (1991).
\end{references}
\end{document}